\documentclass{article}
\usepackage{spconf,amsmath,graphicx}

\usepackage{enumitem}
\setlist{nosep, leftmargin=14pt}

\usepackage{mwe} 
\usepackage{subcaption}
\usepackage{hyperref}
\usepackage{mathtools}
\usepackage{bm}

\title{Improving Out-of-Distribution Detection in Echocardiographic View Classication through Enhancing Semantic Features}
\name{
{\it {Jaeik Jeon$^{1 \star}$ \thanks{$\star$ denotes the first author that contributed equally to this work, while $\dagger$ denotes the corresponding author (jyg1722@gmail.com; yeonyeeyoon@gmail.com).} \qquad 
Seongmin Ha$^{1 \star}$ \qquad 
Yeonggul Jang$^{1 \dagger}$ \qquad
Yeonyee E. Yoon $^{4 \dagger}$ \qquad  
Jiyeon Kim$^{2,3}$}} \\
{\it {Hyunseok Jeong$^{2,3}$ \qquad 
Dawun Jeong $^{2,3}$   \qquad
Youngtaek Hong$^{1,3}$  \qquad 
Seung-Ah Lee $^1$}} \\
{\it Hyuk-Jae Chang$^{1,3,5}$}
}

\address{$^{1}$ Ontact Health, Seoul, South Korea  \\
    $^{2}$Department of Internal Medicine, Graduate School of Medical Science,\\ Brain Korea 21 Project, Yonsei University College of Medicine\\ 
    $^{3}$ CONNECT-AI Research Center, Yonsei University College of Medicine, Seoul, South Korea\\
    $^{4}$Cardiovascular Center, Seoul National University Bundang Hospital, Seongnam, South Korea \\
    $^{5}$Severance Cardiovascular Hospital, Yonsei University Health System, Seoul, South Korea
} 
\begin{document}
%
\maketitle
\begin{abstract}

In echocardiographic view classification, accurately detecting out-of-distribution (OOD) data is essential but challenging, especially given the subtle differences between in-distribution and OOD data. While conventional OOD detection methods, such as Mahalanobis distance (MD) are effective in {\it far}-OOD scenarios with clear distinctions between distributions, they struggle to discern the less obvious variations characteristic of echocardiographic data. In this study, we introduce a novel use of label smoothing to enhance semantic feature representation in echocardiographic images, demonstrating that these enriched semantic features are key for significantly improving {\it near}-OOD instance detection. By  combining label smoothing with MD-based OOD detection, we establish a new benchmark for accuracy in echocardiographic OOD detection.

\end{abstract}


\begin{figure*}[ht!]
\normalsize
\centerline{\includegraphics[width=2.0\columnwidth]{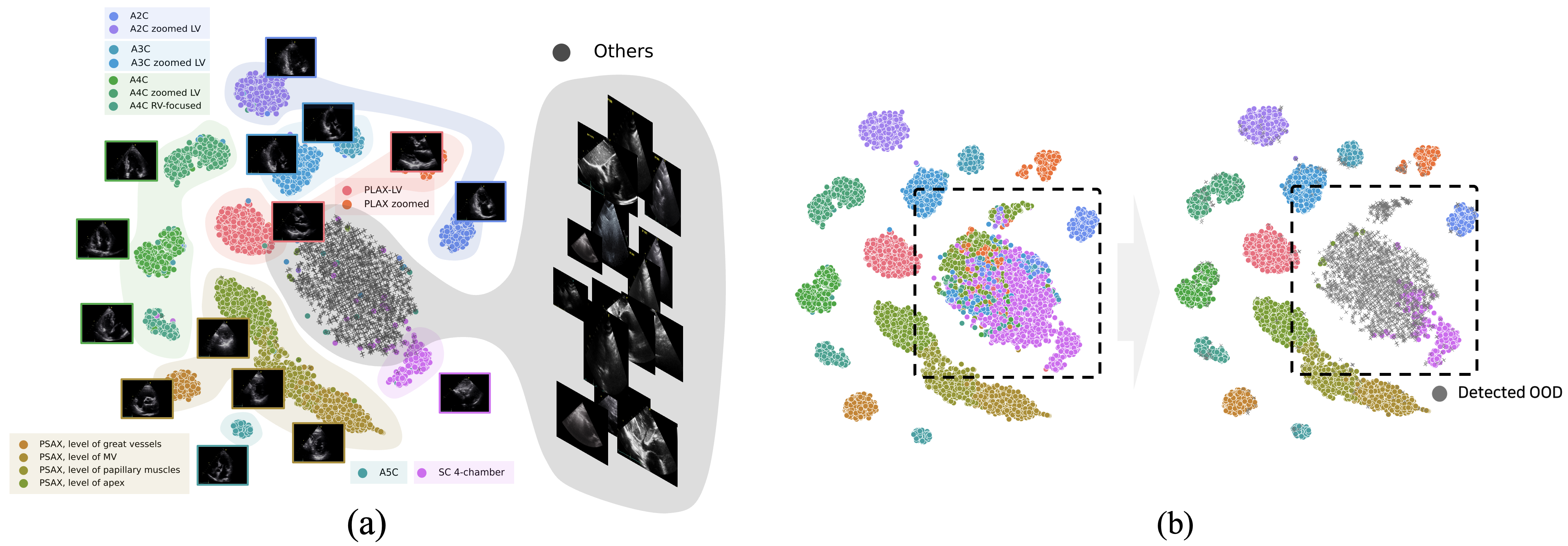}}
\caption{{\bf  Challenges in Echocardiographic Out-of-Distribution (OOD) Detection}
(a) Depicts the similarity in characteristics across standard and OOD echocardiographic views. (b) Demonstrates the model's tendency to classify OOD instances into pre-defined classes without OOD detection, while successfully filtering out them when OOD detection is employed.
 }
\label{fig1}
\end{figure*}

\section{Introduction}

Echocardiography, a primary imaging tool for cardiovascular disease evaluation, captures the heart’s intricate three-dimensional structure through various windows and angles, known as echocardiographic views. Accurate classification of these views is a foundational step for automatic analysis of echocardiography \cite{yoon2021artificial, zhang2018fully}.  However, the wide range of views necessary for classification in echocardiography poses significant challenges.  This field not only includes a broad spectrum of standard views as outlined in guidelines  \cite{mitchell2019guidelines}, but also allows for the acquisition of atypical views, which are influenced by patient-specific conditions and operator expertise.  A major challenge arises in gathering this immense spectrum of both guideline-specified and atypical views for training neural networks. When models are trained on a limited number of views, they misclassify unseen, out-of-distribution (OOD) inputs as one of the predefined classes (Fig. \ref{fig1}). This misclassification lead to errors in subsequent processes such as segmentation of the target structure and automatic measurement of clinical parameters \cite{kusunose2020clinically}. Therefore, there is a pressing need for methods that can effectively detecting and handling these OOD inputs.

Echocardiographic images present a distinct challenge for OOD detection. Unlike conventional research settings, where in-distribution (ID) and OOD data often exhibit pronounced disparities, the differences in echocardiography are far more subtle (Fig. \ref{fig1}). Conventional methods such as Mahalanobis distance (MD) \cite{lee2018simple}, though effective in scenarios with {\it far} OOD samples (e.g. CIFAR-10 vs. SVHN), are less adept at detecting these nuanced differences in echocardiography. This gap in existing methods spurred the development of our approach, which uniquely incorporates label smoothing into MD-based OOD detection. The cornerstone of our approach lies in enhancing of the representation of semantic feature through label smoothing. This technique focuses on the crucial semantic characteristics of echocardiographic images, a crucial aspect for effectively identifying OOD data in a domain where conventional OOD methods have limitations. 

Our methodology aligns with the recent works of emphasizing semantic features in OOD detection, similar to methods such as likelihood ratio \cite{ren2019likelihood} and its adoption in MD \cite{ren2021simple}. 
However, our method differs from these techniques by not requiring a separate {\it background} model to filter out background components; instead, it inherently conditions the model concentrate exclusively on semantic features.
Our approach is a first in the field to build a connection between label smoothing and MD for enhancing semantic features in the context of OOD detection.

\begin{figure*}[ht!]
\normalsize
\centerline{\includegraphics[width=2.0\columnwidth]{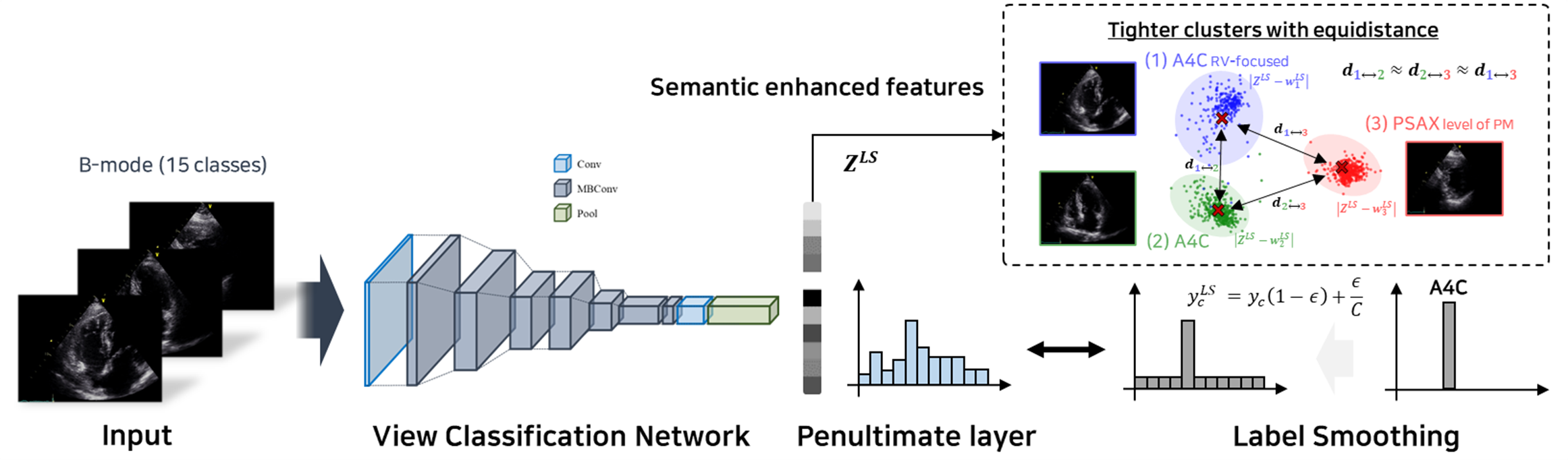}}
\caption{{\bf  Network Training Procedure with Label Smoothing for semantic enhanced features.}
 }
\label{fig: Network}
\end{figure*}

\section{Backgrounds}
\subsection{Out-of-Distribution Detection}
Let $\{(\mathbf x_i, \mathbf y_i)\}_{i=1}^N \coloneqq \mathcal D_{\textrm {in}}$ be an ID dataset. Here, $\mathbf x_i \in \mathcal X$ corresponds to the echocardiographic images and  $\mathbf y_i \in \mathcal{Y}$ is a one-hot encoding vector for class labels $ c \in {1, \dots, C}$. In our study, we identify an OOD sample $(\mathbf x, \mathbf y)$ as one in which the label $\mathbf y$ does not pertain to any of the $C$ ID classes. 



\subsection{Mahalanobis Distance-based OOD Detection}
In OOD detection, the MD-based OOD detector [ref] fits conditional Gaussian distributions $N(\bm \mu_c, \Sigma)$ for each class $c=1,\dots C$ using a feature representation $\mathbf z_i = f(\mathbf x_i)$. A commonly used feature representation is the activation in the penultimate layer of a neural network $f$. The mean vector for each class $c$ and the shared covariance matrix are estimated as $\bm\mu_c=\frac1{N_c}\sum_{i:y_i=c}\mathbf z_i$ and $\Sigma=\frac1N\sum_{c=1}^C\sum_{i:y_i=c}(\mathbf z_i-\bm\mu_c)(\mathbf z_i-\bm\mu_c)^T$, respectively. Then, the MD for a class $c$ for an input $\mathbf x$ is defined as: 
\begin{equation}
    \textrm{MD}_c(\mathbf x) = (\mathbf z - {\bm \mu}_c)^T \hat\Sigma ^{-1} (\mathbf z - {\bm\mu}_c),
\end{equation}
This distance is then used to compute the OOD score for a test input $\mathbf x^{\textrm{test}}$, formulated as $s(\mathbf x^{\textrm{test}}) = -\textrm{min}_c\textrm {MD}_c(\mathbf x^{\textrm{test}})$. This score indicates the degree to which test input deviates from the known class distributions. However, as we will demonstrate in section \ref{experiments}, the MD method exhibits limitations in detecting OOD inputs in echocardiography, a phenomenon also observed in genomic data by \cite{ren2019likelihood}. Specifically, MD struggles in {\it near} OOD environments, where OOD samples closely resemble ID data. 




\subsection{Likelihood Ratio Methods}

Addressing this, \cite{ren2019likelihood} developed a method that emphasizes the semantic component $\mathbf x_S$, comprising distinctive patterns inherent to ID data, and effectively reducing the influence of the background component $\mathbf x_B$, which encapsulates population-level statistics, for an input $\mathbf x=\{\mathbf x_B, \mathbf x_S\}$. 
They computed a likelihood ratio between a foreground model $p_{\bm\theta}(\mathbf x)$ and a background model $p_{\bm\theta_0}(\mathbf x)$, approximated by $\textrm{LLR} (\mathbf x) \approx \log p_{\bm\theta}(\mathbf x_S) - \log p_{\bm\theta_0}(\mathbf x_S)$, to isolate $\mathbf x_S$, and use this as a confidence score for OOD detection. 
\cite{ren2021simple} used the fitted conditional Gaussian distributions $N(\bm \mu_c, \Sigma)$ as the foreground model $p_{\bm\theta}(\mathbf x)$ and a global Gaussian distribution as a proxy for the background model $p_{\bm\theta_0}(\mathbf x)$, represented as $N(\bm\mu_{\textrm{global}}, \Sigma)$, where $\bm\mu_0 = \frac{1}{N}\sum_{i=1}^N \mathbf z_i$. This method effectively translates the Gaussian likelihood into the Mahalanobis Distance, computed as $\textrm{MD}_{c}^{\textrm
{LLR}}(\mathbf x) \coloneqq {\textrm{MD}_\textrm{c}(\mathbf z)} - {\textrm{MD}_{\textrm{global}}(\mathbf z)}$. This is then used as an OOD score $s^{\textrm{LLR}}(\mathbf x^{\textrm{test}}) =-\textrm{min}_c\textrm {MD}^{\textrm{LLR}}_c(\mathbf x^{\textrm{test}})$, offering a semantic-focused metric for OOD detection.

\section{Methods}
\subsection{Label Smoothing}
Our method similarly emphasized the semantic components of the input to effectively detect OOD data. However, diverging from the approaches proposed in previous studies \cite{ren2019likelihood, ren2021simple}, our focus is on enhancing the semantic features in the penultimate layer activations, $\mathbf z_i$.
Central to our approach is the use of label smoothing, which has been widely adopted in various deep networks across multiple tasks as a strategy to prevent overconfidence \cite{szegedy2016rethinking, zoph2018learning}. 
This approach is implemented by adjusting the target distribution $\mathbf y$ for each class label. The modified target distribution $\mathbf y^{LS}$ for a correct class is given by $y_c^{LS} = y_c(1-\epsilon) + \frac\epsilon C$ and $\frac\epsilon C$ for all other classes.

\subsection{Enhanced Semantic Features in Penultimate Layer Activations}
\cite{muller2019does} offered a novel interpretation of label smoothing, suggesting that it encourages the clusters of each classes to be equidistant. This notion is supported by the observations that  the logit $\mathbf z^T\mathbf w_c$ for the last layer weights of the $c^{\textrm{th}}$ class $\mathbf w_c$ can be approximated by the squared Euclidean distance $||\mathbf z-\mathbf w_c||_2^2$. Based on this, \cite{muller2019does} argues that label smoothing might negatively impact the knowledge distillation \cite{hinton2015distilling}, by erasing the relative information between logits. However, we propose a different perspective. We assert that the loss of information is a consequence of prioritizing semantic features which is beneficial for differentiating similar-looking images and especially beneficial for near-OOD detection. Essentially, label smoothing guides the network to focus more on the semantic components in the penultimate layer, resulting in a richer semantic representation, $\mathbf z^{\textrm{LS}} = f^{\textrm{LS}}(\mathbf x) \approx f^{\textrm{LS}}(\mathbf x_S)$.


Leveraging the feature representation trained with label smoothing $\mathbf z^{\textrm{LS}}$ (Fig. \ref{fig: Network}), we compute the $\textrm{MD}_c^{\textrm{LS}}$ for a class $c$:
\begin{align}
\textrm{MD}_c^{\textrm{LS}}(\mathbf x) = (\mathbf z^{\textrm{LS}} - {\bm \mu_c^{\textrm{LS}}})^T \Sigma^{\textrm{LS}^{-1}} (\mathbf z^{\textrm{LS}} - {\bm\mu}_c^{\textrm{LS}})
\end{align}
with
\begin{align}
\bm\mu_c^{\textrm{LS}}&=\frac1{N_c}\sum_{i:y_i=c}\mathbf z_i^{\textrm{LS}}\\ \Sigma^{\textrm{LS}}&=\frac1N\sum_{c=1}^C\sum_{i:y_i=c}(\mathbf z_i^{\textrm{LS}}-\bm\mu_c^{\textrm{LS}})(\mathbf z_i^{\textrm{LS}}-\bm\mu_c^{\textrm{LS}})^T.
\end{align}
This $\textrm{MD}_c^{\textrm{LS}}$ is then used to derive a more semantically sensitive OOD score, $s^{\textrm{LS}}(\mathbf x^{\textrm{test}}) =-\min_c \textrm{MD}^{\textrm{LS}}_c(\mathbf x^{\textrm{test}})$.

\subsection{Dataset and Experimental Setup}
In our study, we employed a dataset from the Open AI Dataset Project (AI-Hub), an initiative of the South Korean government's Ministry of Science and ICT \cite{AI-hub}. Comprising 67,553 B-mode echocardiographic videos across 15 defined categories, these classes encapsulate the most commonly acquired echocardiographic views, especially those requiring quantitative assessments. During the manual classification process by clinical experts, any views not fitting within these predefined categories were assigned to an 'Other' class, designated as OOD (Fig. \ref{fig1}). This 'Other' class, which includes 2,802 videos not matching the standard 15 but still clinically significant, provided a unique opportunity to evaluate the effectiveness of our enhanced semantic feature-based OOD detection method.

\section{Experiments}
\label{experiments}


\subsection{Implementation Details}
For classification network, we employed Efficientnet-B3 model \cite{tan2019efficientnet} with images resized to 224 × 224 pixels and normalized within [-1,1]. 
All frames in the B-mode video were used for training, only the first frame was used for evaluating classification performance. 
During training, we used all frames from B-mode videos, but for evaluation, we averaged frame predictions to assess performance at the video level.
The cross-entropy loss was optimized using the Adam optimizer \cite{kingma2014adam} at a learning rate of 0.001 and cosine annealing learning rate scheduling \cite{loshchilov2016sgdr}. 
In experiments where label smoothing was applied, $\epsilon=0.1$ was used.
The model was trained over 200 epochs, selecting the best performer based on validation loss, and RandAug data augmentation \cite{cubuk2020randaugment} was employed for enhanced regularization. We also employed a 10-fold cross-validation approach for comprehensive evaluation.

\noindent
\begin{figure}[t!]
  \centering
    \includegraphics[width=0.44\textwidth]{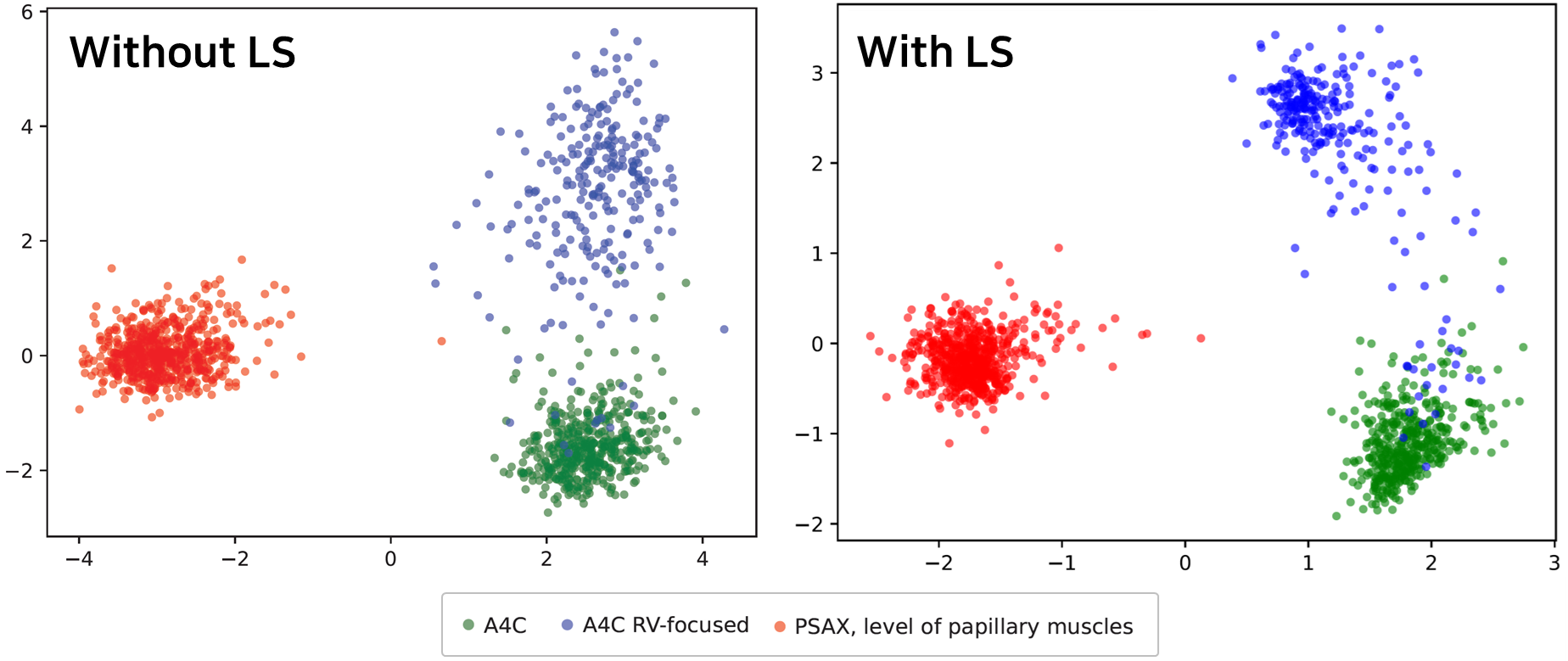}
\captionof{figure}{{ \bf{Visualizations of Penultimate Layer Activations in the Test Set}}}
\label{fig3}
\end{figure}

\begin{figure}[t!]
  \centering
    \includegraphics[width=0.38\textwidth]{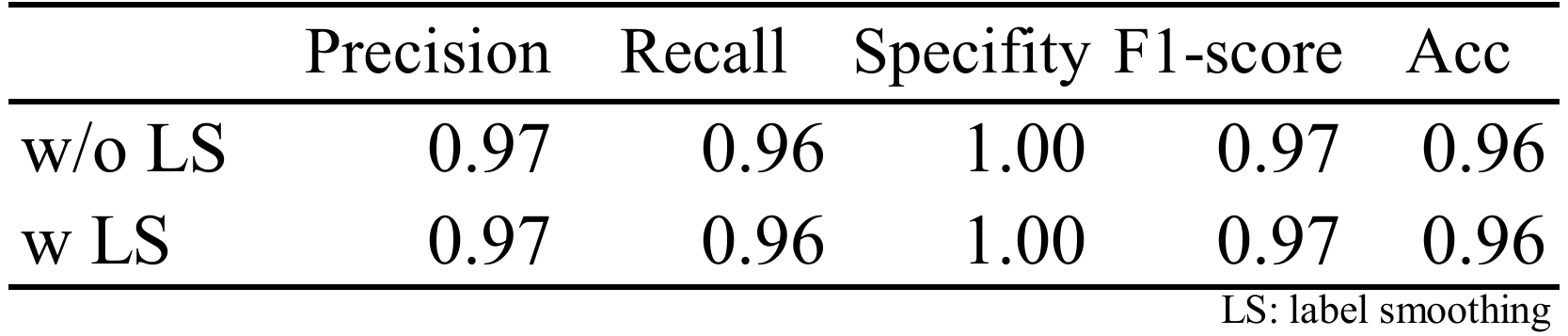}
\captionof{table}{{ \bf{Classification Performance in In-Distribution}}}
\label{table 2}
\end{figure}

\subsection{Penultimate Layer Activations Visulaization}

To investigate the impact of label smoothing on penultimate layer activations, we employed a visualization method inspired by \cite{muller2019does} in Fig. \ref{fig3}. 
First, we chose three echocardiographic view classes: apical 4 chamber (A4C), A4C right venvticular (RV)-focused (A4C RV-focused), and parasternal short axis (PSAX) view.  While PSAX view is markedly different from the other two, A4C and A4C RV-focused views fundamentally share characteristics of the A4C view, with the main difference being whether the focus is on the left ventricle (LV) or the right ventricle (RV). We then identified an orthonormal basis for a plane that intersects their weights, and projected the penultimate layer activations of samples from these classes onto this plane.
This reveals that without label smoothing, the clusters for A4C and A4C RV-focused are relatively close together, suggesting that their representations include components of the background element $\mathbf x_B$ common to both views. However, with the application of label smoothing, each cluster becomes more distinct and tighter, indicating an increase in equidistance. This transformation implies that the feature representations are now more distinguishable between A4C and A4C RV-focused, focusing more on the semantic component $\mathbf x_S$ that differentiates them.

In contrast to the advantages of label smoothing in enhancing semantic feature representation, Table \ref{table 2} shows that its application does not necessarily translate into improved classification performance. This finding diverges from previous studies reporting enhancements in classification accuracy through label smoothing \cite{szegedy2016rethinking, zoph2018learning}, highlighting its specific utility in semantic distinction rather than classification performance improvement.


\noindent
\begin{figure}[t!]
  \centering
    \includegraphics[width=0.39\textwidth]{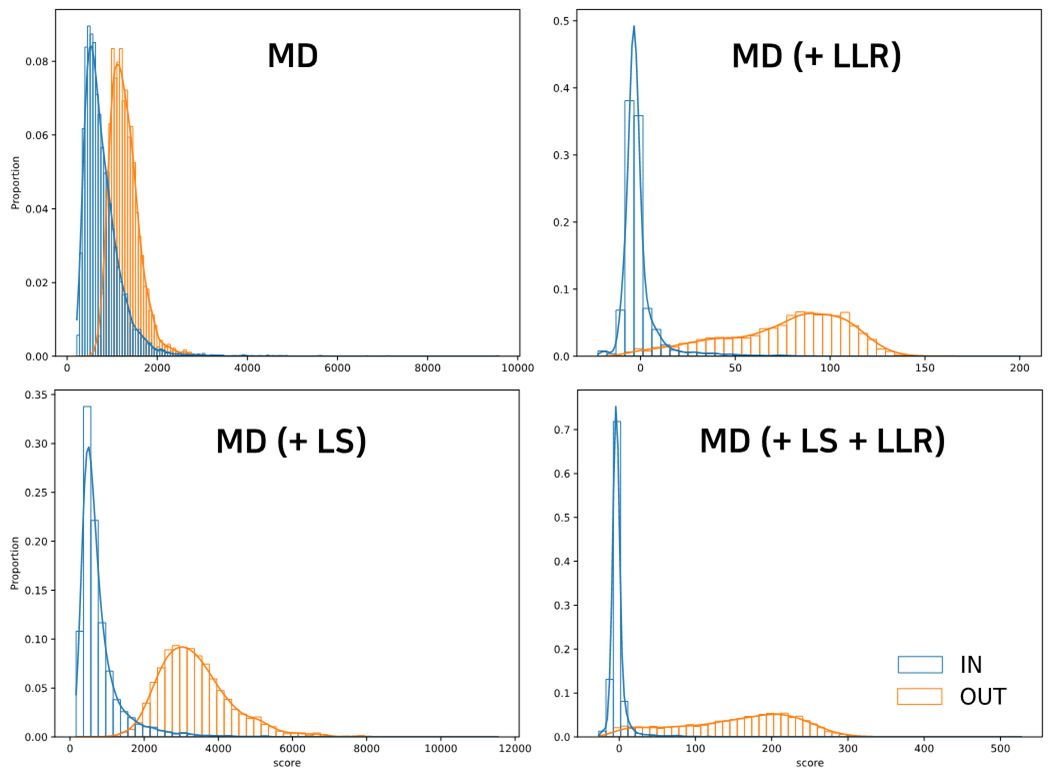}
\captionof{figure}{{ \bf{Out-of-Distribution Detection Score Density Plot}}}
\label{fig 4}
\end{figure}
\noindent
\begin{figure}[t!]
  \centering
    \includegraphics[width=0.45\textwidth]{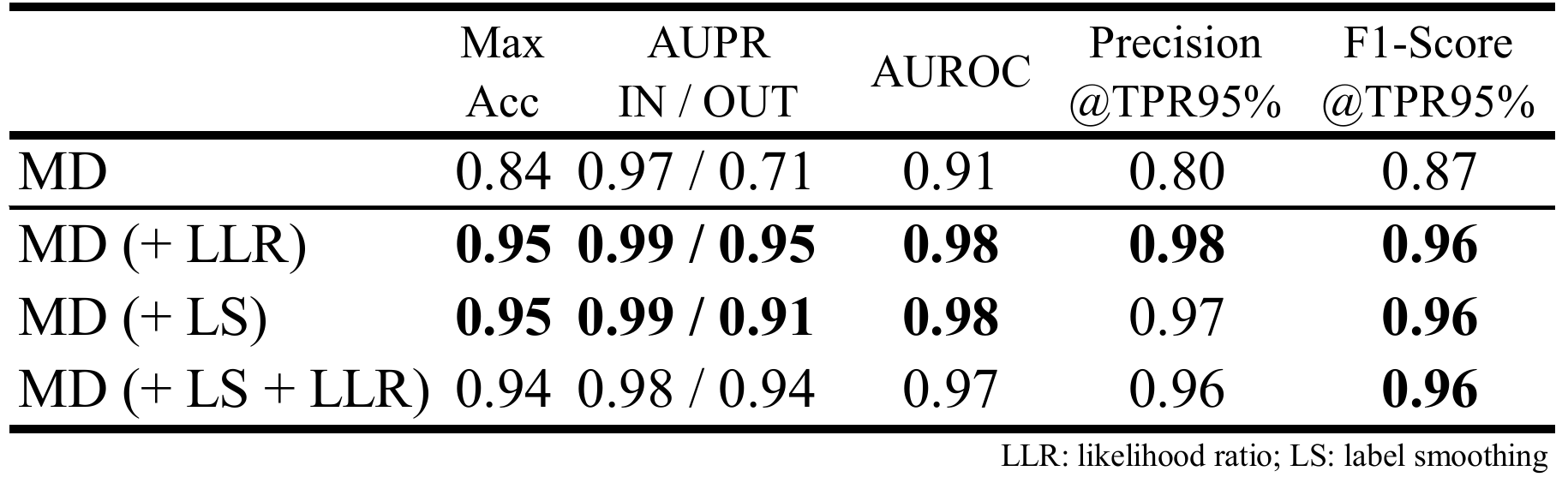}
\captionof{table}{{ \bf{Out-of-Distribution Detection Performance}}}
\label{table 1}
\end{figure}

\subsection{Out-of-Distribution Detection Performance}

We employed a range of metrics drawn from established literature  \cite{hendrycks2016baseline, lee2018simple} to evaluate our proposed OOD detection performance. These include area under the receiver operating characteristic curve (AUROC), and area under the precision-recall curve (AUPR). We report AUPR in two forms: AUPR-In, which focuses on identifying ID samples, and AUPR-Out, aimed at correctly identifying OOD samples. Additionally, we provide precision and F1-scores calculated at a threshold yielding a true positive rate (TPR) of 0.95. 

Our analysis starts with Fig. \ref{fig 4}, showcasing a density plot of OOD scores. This highlights the improved separability of scores between ID and OOD instances, achieved through the application of semantic-enhanced methods.
Table \ref{table 1} supports this with quantitatively analysis showing that semantic-focused methods, $\textrm{MD}^{\textrm{LLR}}$ and $\textrm{MD}^{\textrm{LS}}$, surpass the baseline $\textrm{MD}$ approach, achieving a 95\% success rate compared to the baseline's 84\%. 
Interestingly, the combination of label smoothing with the likelihood ratio method results in a slight decrease in performance. 
This effect can be attributed to the diminished background component in $\mathbf{z}^{\textrm{LS}}$. This contradicts the key assumption of the likelihood ratio method, which posits that both models should capture background information equally well. Thus, this result serves as additional evidence that label smoothing successfully erase background information, enhancing the focus on semantic components in the echocardiographic images.




\section{Conclusion}
Our research demonstrates label smoothing's effectiveness in MD-based OOD detection, notably in identifying echocardiographic near-OOD instances. It establishes a novel link between label smoothing and MD, uniquely eliminating the need for a background model by focusing the model exclusively on semantic features, in contrast to likelihood ratio methods.

\newpage

\section{Compliance with ethical standards}
\label{sec:ethics}
This research study was conducted using the OpenAI Dataset Project (AI-Hub) collected from five tertiary hospitals, and the institutional review board of each hospital approved the use of de-identified data and waived the requirement for informed consent owing to the retrospective and observational study design (IRB No. 2021-0147- 003; CNUH 2021-04-032; HYUH 2021-03-026-003; SCHBC 2021-03-007-001; B-2104/677-004).

\section{Acknowledgments}
\label{sec:acknowledgments}

This work was supported by the Institute of Information \& communications Technology Planning \& Evaluation (IITP) grant funded by the Korean government (MSIT) (No.
\

\noindent
$2022000972$, development of flexible  mobile health care softwar eplatform using 5G MEC).

\bibliographystyle{IEEEbib}
\bibliography{strings,refs}

\end{document}